\documentclass[twocolumn,showpacs,preprintnumbers,amsmath,amssymb]{revtex4}

\usepackage{epsf}
\usepackage{graphicx}  
\usepackage{dcolumn}   
\usepackage{bm}        

\newcommand{\be}{\begin{equation}}
\newcommand{\en}{\end{equation}}
\newcommand{\bea}{\begin{eqnarray}}
\newcommand{\ena}{\end{eqnarray}}
\newcommand{\hbo}{\hbox to 1 true cm {\hfill } }
\newcommand{\Dcal}{{\mathcal D}}
\newcommand{\Lcal}{{\mathcal L}}
\newcommand{\Det}{{\rm Det}}
\newcommand{\vD}{{\bf D}}
\newcommand{\vA}{{\bf A}}
\newcommand{\vPi}{{\bf \Pi}}
\newcommand{\vB}{{\bf B}}
\newcommand{\vJ}{{\bf J}}

\newcommand{\vx}{{\bf x}}
\newcommand{\vy}{{\bf y}}
\newcommand{\vp}{{\bf p}}

\begin{document}

\preprint{UNITUE-THEP/5-2004}

\title{ Propagators in Coulomb gauge from SU(2) lattice gauge theory }

\author{Kurt Langfeld}
\author{Laurent Moyaerts}

\affiliation{
Insitut f\"ur Theoretische Physik, Universit\"at T\"ubingen\\ 
D-72076 T\"ubingen, Germany. 
}

\date{ June 15, 2004 }

\begin{abstract}
A thorough study of 4-dimensional SU(2) Yang-Mills theory in Coulomb gauge 
is performed using large scale lattice simulations. The 
(equal-time) transverse gluon propagator, the ghost form factor $d(p)$ and 
the Coulomb potential $V_{coul} (p) \propto d^2(p) f(p)/p^2$ 
are calculated. For large momenta $p$, the 
gluon propagator decreases like $1/p^{1+\eta}$ with $\eta =0.5(1)$. 
At low momentum, the propagator is weakly momentum dependent. 
The small momentum behavior of the Coulomb potential is consistent 
with linear confinement. We find that the inequality $\sigma _{coul} \ge
\, \sigma $ comes close to be saturated. 
Finally, we provide evidence that the ghost form factor $d(p)$ and $f(p)$ 
acquire IR singularities, i.e., $d(p) \propto 1/\sqrt{p}$ and 
$f(p) \propto 1/p$, respectively. It turns out that the combination 
$g_0^2 d_0(p)$ of the bare gauge coupling $g_0$ and the bare ghost 
form factor $d_0(p)$ is finite and therefore renormalization group invariant. 

\end{abstract}


\maketitle

\section{Introduction} 

Quark confinement is the key property which dictates the structure of 
matter at low temperatures and densities. It is 
attributed to the low energy regime of Quantum Chromo Dynamics (QCD) 
and is, therefore, only accessible by non-perturbative techniques. 
Lattice gauge theory provides a gauge invariant regularization 
of Yang-Mills theory, and corresponding numerical simulations 
do not require gauge fixing. 

\vskip 0.3cm 
More than twenty years ago, Mandelstam and 't~Hooft pointed out that 
gauge fixing might be convenient for identifying the degrees of freedom 
which are relevant for confinement. Over the recent past, evidence has been 
accumulated by means of lattice simulations that topological obstructions 
of the gauge field, such as monopoles and vortices,  are responsible for 
confinement (see~\cite{Greensite:2003bk} for a recent review). 
Non-perturbative approaches based upon Dyson-Schwinger equations 
(DSE)~\cite{Roberts:2000aa,Alkofer:2000wg}, variational 
techniques~\cite{Szczepaniak:2001rg,Szczepaniak:2003ve,Feuchter:2004gb} 
and flow equations~\cite{Jungnickel:mn,Pawlowski:2003hq} necessarily 
involve gauge fixing. These approaches address QCD Green functions which 
encode information on confinement at low momenta. 

\vskip 0.3cm 
In Landau gauge, it was firstly put forward  by Gribov~\cite{Gribov:1977wm}
and further elaborated by Zwanziger~\cite{Zwanziger:1991ac} that 
quark confinement is associated with a divergence of the ghost form factor 
in the infrared limit. First indications that the so-called 
horizon criterion for confinement is satisfied were 
reported in~\cite{Suman:1995zg}. Using truncated DSEs, strong 
evidence for an IR divergent ghost form factor was reported 
in~\cite{vonSmekal:1997is,vonSmekal:1997is2}. Subsequently, many efforts 
were devoted to detect the low energy behavior of the ghost form factor by 
means of analytic~\cite{Zwanziger:2001kw,Bloch:2001wz,Lerche:2002ep, 
Fischer:2002hn,Alkofer:2003jj,Bloch:2003yu} and lattice~\cite{Cucchieri:1997dx,
Bakeev:2003rr,Bloch:2003sk} techniques. A good qualitative agreement 
between the DSE and lattice results was found. Moreover, a tight 
relation between the vortex picture of confinement and the 
Gribov-Zwanziger criterion was given in~\cite{Gattnar:2004bf}. 

\vskip 0.3cm 
Coulomb gauge is most convenient for a variational approach to 
Yang-Mills Green functions~\cite{Szczepaniak:2001rg,Szczepaniak:2003ve,
Feuchter:2004gb}. The corresponding Hamiltonian contains an 
instantaneous interaction between color sources, the so-called 
Coulomb potential $V_{coul}(r)$. Hence, Coulomb gauge QCD offers 
the possibility to address the confining potential by purely 
analytic considerations. It has been shown 
recently~\cite{Zwanziger:2002sh} that the Coulomb potential 
is an upper bound of the static quark-antiquark potential. In pure 
Yang-Mills theory, both potentials are expected to be linearly rising at 
large distances $r$. At present, lattice simulations provide two 
different values for the  asymptotic Coulomb force: A calculation 
of the potential involving the zeroth component of the gauge 
field yields $\sigma _{coul} \approx (2 \, - \, 3) \, 
\sigma $~\cite{Greensite:2003xf,Greensite:2004ke}, while 
 $\sigma _{coul} \approx \sigma $~\cite{Cucchieri:2002su} is reported 
if the Green function which defines the Coulomb potential
is directly evaluated.

\vskip 0.3cm 
The variational approach invokes a quasiparticle picture for the 
(transverse) gluons. The trial wave functional generically 
is Gaussian, and the gluon dynamics is encoded in the (quasi particle) 
gap function $\omega _{QP}$ (which is derived in a 
gluonic quasiparticle picture from the inverse of the 
instantaneous gluon propagator). It was recently proposed 
to supplement the Faddeev-Popov determinant (see below) to the 
wave functional in order to strengthen the impact of configurations 
close to the Gribov horizon~\cite{Feuchter:2004gb}. 
The latter modification of the wave functional has certainly a 
strong impact on the gap function at least at small momenta: 
while it is suggested in~\cite{Szczepaniak:2001rg} that $\omega  _{QP}$ 
approaches a constant in the IR limit, an IR 
divergence was reported in~\cite{Feuchter:2004gb}. 
Both approaches employ a truncation of the resulting Dyson equations  
which does not account for wave functional renormalization. 

\vskip 0.3cm 
In the present paper, we perform a thorough lattice investigation 
of the transverse gluon propagator, the ghost form factor and 
the Coulomb potential. Our results suggest that (at least in 
four dimensions) the equal time gluon propagator 
is weakly momentum dependent in the IR regime. 
For the first time, wave 
function renormalization constants for the gluon and ghost 
fields are obtained. Our lattice results indicate quite a sizable 
anomalous dimension for the gluon propagator yielding 
$G (\vert \vp \vert ) \rightarrow 1/\vert \vp \vert ^{1 + \eta}$ 
where $\eta \approx 0.5 $. Our results for the Coulomb potential 
are compatible with linear confinement and suggest that 
the inequality $\sigma _{coul } \ge \sigma $ is almost saturated.  
This finding favors the result in~\cite{Cucchieri:2002su}. 
We will find, however, that the result $\sigma _{coul } \approx (2-3) \, 
\sigma $, reported in~\cite{Greensite:2003xf,Greensite:2004ke}, 
cannot be ruled out.

\section{ \label{sec:parf} The partition function }

Let us briefly review the relation between the functional integral 
formulation and the Hamilton formulation of Coulomb gauge Yang-Mills 
theory following~\cite{Cucchieri:2000hv}. 
Our starting point will be the generating functional for Euclidean 
Green functions in Coulomb gauge, i.e., 
\bea  
Z [J] &=& \int \Dcal A \exp \left\{ \int \Lcal dx \right\} 
\delta(\partial _i A^i_a) \; \Det[-\nabla\cdot \vD], 
\label{eq:2} \\ 
\Lcal &=& - \frac{1}{4} F_{\mu\nu}^a F^{\mu\nu}_a \; - \; i \,  g_0  \; 
A_\mu^a J^\mu_a \; ,
\nonumber 
\ena 
where $A_\mu ^a$ are the gauge fields, $F_{\mu\nu}^a$ denotes the usual field 
strength tensor and $J^\mu_a $ is an external source. \hfill \break
$ \Det[-\nabla\cdot \vD]$ is the Faddeev-Popov determinant ($\vD $ 
is the gauge covariant derivative), and $g_0$ 
the bare gauge coupling. Equation (\ref{eq:2}) will serve as the 
starting point for the lattice simulations reported in this work. 

\vskip 0.3cm  
The canonical momenta conjugated to the gauge fields are introduced by 
$$
\int \Dcal \vPi \; 
\exp{\left\{-  \int dx \; \left( \frac{1}{2}\vPi^2 \; - \;  i \; 
\Pi_i^a F^{0i}_a \right) \right\}} 
\; = 
$$
$$
\exp{\left\{- \frac{1}{2}  \int dx \; F_{0i}^aF^{0i}_a \right\}}
\; .  $$
Eliminating the non-dynamical part of the gauge fields,  
$A_0^a$, implements Gauss's law in the partition function, i.e., 
\begin{widetext} 
\be 
 Z=\int \Dcal \vA \; \Dcal \vPi \; \exp \biggl\{ \int \Bigl[ i \; \Pi_i^a
  \dot{A}^{i,a}-\frac{1}{2}(\vPi^2+\vB^2)-g_0 \vA^a \cdot\vJ^a\Bigr] 
\biggr\} \; 
\delta \Bigl( \partial _i A^a_i \Bigr) \; 
\delta \Bigl([D_i\Pi^i]^a-g_0J_0^a \Bigr) \; 
\Det[-\nabla\cdot\vD]. 
\label{eq:3}
\en 
\end{widetext}
Let us introduce the transverse and longitudinal parts of the 
canonical momenta by $\vPi^a=\vPi^a_\perp-\nabla \phi^a$.  
The integration measure factorizes, i.e., 
$\Dcal \vPi\simeq \Dcal \vPi_\perp \Dcal \phi$, 
and the integration over $\phi$ cancels the Faddeev-Popov determinant. 
One finally obtains 
\be 
Z=\int \Dcal \vA_\perp \Dcal \vPi_\perp \exp{\left\{ \int
 (i \, \Pi_{\perp,i}^a\dot{A}^{i,a}-\mathcal{H}_\perp)\right\}} \; . 
\label{eq:4}
\en
Thereby, the Hamiltonian $\mathcal{H}_\perp$ is given by 
\bea
\mathcal{H}_\perp  &=& \int d^3x
\left[\frac{1}{2}(\vPi^2_\perp+\vB^2)+g_0\vA_\perp\cdot\vJ\right] 
\nonumber \\ 
&-& \frac{1}{2 C_2} \; \int \; d^3x \; d^3y \; 
\rho(\vx) \;  \mathcal{V}_{\rm Coul}(\vx,\vy) \;  \rho(\vy) , 
\label{eq:5}
\ena 
where $C_2$ is the quadratic Casimir of the gauge group SU(2), i.e., 
$C_2=3/4$, and $\vB$ is the chromo-magnetic field. 
We have introduced the color charge density by 
\be 
\rho (x) \; = \; f^{abc} A_i^b(x)  \Pi_{\perp \, i} ^c(x) + J_0(x) \; . 
\en 
The ``tree level'' term which mediates the interaction between 
color charges, i.e., 
\be
\mathcal{V}_{\rm Coul}(\vx,\vy) \; = - C_2 \; g_0^2 
\; M^{-1}(-\Delta)M^{-1}|_{(\vx,\vy)} 
\; , 
\label{eq:6}
\en 
will give rise to the so-called  Coulomb potential upon averaging 
over the gauge fields. Thereby, 
\be 
M=-\nabla\cdot\vD 
\label{eq:fpm}
\en 
is the Faddeev-Popov matrix. 

\section{The numerical approach } 

\subsection{ Setup } 

Configurations of a $N^4$ cubic lattice with lattice spacing $a$ 
are generated using the standard Wilson action. 
A lattice configuration is represented by a set of unitary matrices 
$U_\mu (x) \; \in \; SU(2) $. The size of the lattice spacing 
as a function of the Wilson $\beta $ parameter is obtained from 
the interpolating formula~\cite{Bloch:2003sk}
\be 
\ln \left( \sigma a^2 \right) \; = \; - \; 
\frac{4 \pi^2}{\beta_0} \, \beta \; + \; 
\frac{2 \beta_1}{\beta_0^2} \, \ln \left( \frac{4 \pi^2}{
\beta_0} \, \beta \right) \; + \; \frac{4 \pi^2}{\beta_0} \, 
\frac{d}{\beta} \; + \; c \; , 
\label{eq:tension}
\en 
where 
\be
\beta_0 \; = \; \frac{22}{3} \; ,  \hbo
\beta_1 \; = \; \frac{68}{3}  \; . 
\label{eq:b0b1}
\en 
The first two terms on the r.h.s.\ of (\ref{eq:tension}) are in 
accordance with 2-loop perturbation theory. The term $ d / \beta $ 
represents higher-order effects and the term $\,c\,$
is a dimensionless scale factor to the string tension. 
It was observed~\cite{Bloch:2003sk} that the choice 
\be 
c \; = \; 4.38(9) \; , \hbo  d \; = \; 1.66(4) 
\label{fitpara} 
\en 
reproduces the measured value $\sigma a^2$ to very good precision. 

\vskip 0.3cm 
Two possible definitions of the gauge field $A_\mu ^b (x)$ 
in terms of the link matrices are explored in the present paper. 
Decomposing a particular SU(2) matrix by 
\be 
U_\mu (x) \; = \; u^0_\mu (x) \; + \; i \, u^a_\mu (x) \; \tau ^a \; , 
\en 
where $\tau ^a$ are the Pauli matrices, the standard definition 
of the gauge field is given by 
\be
 a\, g_0 \, A^b_{\mu}(x) \; + \; {\cal O}(a^3) 
\; = \; 2 \, u^b_\mu(x) \; . 
\label{eq:fund}
\en
Noticing that the continuum gauge fields actually transform 
under the adjoint representation, the definition 
\be
a\, g_0 \, A^b_{\mu}(x) \; + \; {\cal O}(a^3) 
\; = \; 2 \, u^0_\mu(x) \, u^b_\mu(x) 
\label{eq:tada}
\en
accounts for this particular transformation 
property~\cite{Langfeld:2001cz}. Note that both definitions coincide 
in the continuum limit $a \rightarrow 0$. In the context of minimal 
Landau gauge, both definitions also coincided at the practical 
level within the scaling window~\cite{Bloch:2003sk}. 

\vskip 0.3cm 
Finally, we define the momentum on the lattice 
for the particular direction $\mu $ 
\be 
p_\mu \; := \; \frac{2}{a} \; \sin \Bigl( \frac{ \pi }{N} n_\mu \Bigr) \; , 
\en 
where $-N/2 < n_\mu \le N/2 $ labels the Matsubara frequency. 
This definition minimizes rotational symmetry breaking for 
momenta close to the boundary of the Brillouin zone.

\subsection{Propagators and Coulomb potential }
 
Once and for all, we choose a given time slice of the four dimensional 
space-time by fixing $t=t_0$, and consider propagators which are defined 
within the emerging 3-dimensional hypercube. 

\vskip 0.3cm 
The (bare) gluon propagator is defined by
\be
 G_{ij}^{0 \; ab}(\vx-\vy) \; := \; \langle A_i^a(\vx,t_0) \; 
A_j^b(\vy,t_0) \rangle. 
\label{eq:gluon}
\en
The corresponding propagator in momentum space, i.e., 
$$
 G_{ij}^{0 \; ab}(\vp)=\int d^3 x\  G_{ij}^{0 \; ab}(\vx)\, e^{i\vp\cdot\vx}, 
$$ 
is transversal by virtue of the gauge condition $\partial _i 
A^i_a (\vx, t_0) =0$ and diagonal in color space, i.e., 
\bea
G_{ij}^{0 \; ab}(\vp) &=& \left(\delta_{ij}-\frac{p_ip_j}{\vp^2}\right) 
\; \delta ^{ab} \; G^0(\vp) \; ,  
\\
G^0(\vp) &=& \frac{ f^{0}_g(\vp) }{\vert \vp \vert } \; . 
\label{eq:gluonm}
\ena 
The dimensionless quantity $f^{0}_g$ is the (bare) gluon form factor. 
Since the gauge potential has energy dimension one, 
the quantity $G^0(\vp)$ has the dimension of an 
inverse energy.  

\vskip 0.3cm 
The (bare) ghost propagator is defined as the expectation value of the inverse 
Faddeev-Popov operator $M$ in (\ref{eq:fpm}), i.e., 
\be 
D^{0 \; ab}(\vx-\vy) = \left\langle M^{-1}[A]|^{ab}_{(\vx,\vy)}\right\rangle.
\label{eq:ghostp} 
\en 
Since $D^{ab}$ is diagonal in color space, we may write 
$$ 
D^{0 \; ab}(\vp)=\delta^{ab}D^0(\vp), \hbo  D^0(\vp)=
\frac{d^{0} (\vp)}{|\vp|^2} 
$$ 
in momentum space. 
Thereby, we have introduced the (bare) {\it ghost form factor} 
$d^{0}(\vp)$. 
Since the quantity $\left\langle M^{-1}[A]|_{(\vx,\vy)} \right\rangle$ has 
energy dimension $1$, the ghost form factor $d^{0}(\vp)$ is 
dimensionless and reduces to unity for a free theory. 

\vskip 0.3cm 
The Coulomb potential is given by the expectation value of the 
expression (\ref{eq:6}): 
\begin{align} 
 V^{0}_{\rm Coul}&(\vx-\vy)\delta^{ab} \; = \; 
\nonumber \\ 
-&\, C_2\; g_0^2 \left\langle \left[M^{-1}[A](-\Delta)
M^{-1}[A]\right]_{(\vx,\vy)}^{ab}\right\rangle \; . 
\label{lat:vcoul}
\end{align} 
For the Fourier transform of the Coulomb potential, we make 
the ansatz 
\be 
V^{0}_{\rm Coul}(\vp)=-C_2 \, g_0^2\,\frac{d_{0}^2(\vp)
f(\vp)}{|\vp|^2}
\label{eq:coul}
\en 
where $d^{0}(\vp)$ is the bare ghost form factor. 
The dimensionless function $f(\vp)$
measures the deviation from the factorization 
\begin{align} 
\Bigl\langle M^{-1}[A] \; &(-\Delta)
M^{-1}[A] \Bigr\rangle \; = \; 
\nonumber \\ 
& \Bigl\langle M^{-1}[A] \Bigr\rangle \; (-\Delta) \; 
\Bigl\langle M^{-1}[A] \Bigr\rangle \; , 
\label{eq:fa}
\end{align} 
in which case $f(\vp)=1$. 

\subsection{Renormalization} 
\label{renorm}

Renormalization of Yang-Mills theories in four dimensions implies
that the bare coupling acquires a dependence on the
ultraviolet (UV) cutoff $\Lambda_{UV}$, i.e., 
\be 
g_0 \rightarrow g_0( \Lambda_{UV}/ \Lambda_{scale}) \, . 
\en 
Rather than the bare coupling constant, 
the Yang-Mills scale parameter $ \Lambda_{scale} $ plays 
the role of the only parameter of the theory. 
In the context of (quenched) lattice gauge simulations the string 
tension $\sigma $ is widely used as the generic low-energy scale. 
In this case, the cutoff dependence of the bare 
coupling is implicitly given 
by the $\beta $ dependence of $\sigma \, a^2( \beta ) $ 
where $\beta = 4 /g_0^2$ is related to the cutoff by 
$\Lambda _{UV} = \pi/a(\beta )$. Finally, the 
relation between $\sigma a^2$ and $\beta $ is provided 
by the formula (\ref{eq:tension}) which interpolates 
the calculated values. 

\vskip 0.3cm
In addition, wave-function renormalization constants generically 
develop a dependence on $\Lambda_{UV}/ \Lambda_{scale}$. 
The lattice bare form factors of the previous subsections,
$f^{0}_g$ and $d^{0}$, 
depend on the momentum $p^2$ 
and on the UV cutoff $\Lambda_{UV}$ (given in units of the
string tension) or, equivalently, on the lattice coupling $\beta$. 
The renormalized form factors are obtained upon multiplicative 
renormalization
\bea 
f^\mathrm{ren}_g (p^2, \mu^2) &=& Z_3^{-1}(\beta, \mu) \; 
f^{0}_g(p^2, \beta) 
\label{eq:FR} \\
d^\mathrm{ren} (p^2, \mu^2) &=& \widetilde{Z}_3^{-1}(\beta, \mu) 
\; d^{0}(p^2, \beta) 
\label{eq:JR}
\ena 
using the renormalization conditions 
\be 
 f^\mathrm{ren}_g (\mu^2, \mu ^2)  \; = \; 1 \; , \hbo 
 d^\mathrm{ren} (\mu^2, \mu ^2)  \; = \; 1 \; . 
\label{eq:renorcond} 
\en 
Finally note that the Coulomb potential $V_{\rm Coul}(\vp)$ (\ref{eq:coul}) 
appears as a convolution of propagators. Let us factor out the 
wave function  renormalization of the ghost fields, and let us 
introduce the renormalization constant $Z_f$ of the 4-point vertex 
function involving four ghost fields: 
\be 
V^\mathrm{ren} _{\rm Coul}(\vp, \mu ^2) \; = \; Z_f^{-1}(\beta, \mu) \; 
\widetilde{Z}_3^{-2}(\beta, \mu) V^\mathrm{0} _{\rm Coul}(\vp, \beta ) 
\; , 
\label{eq:vcren} 
\en 
where $ V^\mathrm{0} _{\rm Coul}$ is given in (\ref{lat:vcoul}). 
If the theory is renormalizable without a 4-ghost counter term at tree level, 
then 
\be 
\lim _{\beta \to \infty}  Z_f(\beta, \mu) \; = \; \mathrm{constant} \; , 
\en 
and the ``factorization'' function $f(\vp)$, which is 
implicitly defined by (\ref{eq:coul}), does not acquire an UV divergence. 
Below, we will suppress the superscript ``ren'' and only deal with 
renormalized quantities. 

\vskip 0.3cm 
Multiplicative renormalizability of the theory 
implies that a rescaling of the data for each $ \beta $ value (independently
of the momentum) is sufficient to let 
the form factors  fall on top of a single curve describing the momentum 
dependence of the corresponding renormalized quantity. 
In practice, suitable ``matching factors'' are determined which 
``collapse'' data obtained at different $\beta$ on a single 
curve. The matching factors are then directly related to 
the wave function renormalization constants. This procedure 
is described in detail in ~\cite[Sec.\ V.B.2]{Leinweber:1998uu}
and~\cite{Bloch:2003sk}.

 \subsection{Gauge fixing } 

Let us consider the time slice $t_0$ in which we define 
the transverse (equal time) gluon propagator (\ref{eq:gluon}) 
and the ghost propagator, respectively. 

The lattice configurations are generated without any preference for 
a particular gauge. Subsequently, the gauge transformations $\Omega (x)$ 
are adjusted according to 
\be 
F_U[\Omega ] \; = \; \sum_{\vec{x},i=1 \ldots 3} 
{\rm Re\,Tr}\Bigl(1-U_i^\Omega (\vec{x},t_0) \Bigr)  
\stackrel{\Omega }{\longrightarrow } \mathrm{min} \; , 
\label{eq:gcond}
\en
where 
$$ 
U_\mu ^\Omega (x) \; = \; \Omega (x) \; U_\mu (x) \; \Omega ^\dagger 
(x +\mu) \; . 
$$
It is well known that the condition (\ref{eq:gcond}) fixes the gauge 
up to gauge transformations which only depend on time, i.e., 
$\Omega (t)$. Note, however, that the average over the unfixed 
gauge degree of freedom does not affect the (equal time) gluon 
propagator. The same is true for the ghost propagator. 

\vskip 0.3cm 
The calculation of the non-instantaneous gluon propagator, 
i.e., 
\be
 G^{\mathrm{gen}\, ab}_{ij}(\vx-\vy,t-t_0) \; := \; \langle A_i^a(\vx,t) \; 
A_j^b(\vy,t_0) \rangle 
\label{eq:gluong}
\en
would  inevitably require complete gauge fixing, since the residual gauge 
freedom would imply 
$$
 G^{\mathrm{gen}\, ab}_{ij}(\vx-\vy,t-t_0) \; = \; 0 \; \hbo 
\mathrm{for} \; t\not= t_0 \; . 
$$
The standard procedure is to involve the gauge field $A^a_0(x)$ 
for residual gauge fixing. An example is the Cucchieri-Zwanziger 
condition~\cite{Cucchieri:2000gu}
\be 
F_t[\Omega ] \; = \; \sum_{\vec{x},t} 
{\rm Re\,Tr}\Bigl(1-U_0^\Omega (\vec{x},t) \Bigr)  
\stackrel{\Omega (t) }{\longrightarrow } \mathrm{min} \; . 
\label{eq:gcond2}
\en
It is also possible to fix the residual gauge freedom without 
incorporating  $A^a_0(x)$. In this case, the  $A^a_0(x)$ 
integration in section \ref{sec:parf} can still be performed 
thereby enforcing Gauss's law (see (\ref{eq:3})). The advantage is that 
the tight relation between the functional approach (\ref{eq:2}) and 
the Hamiltonian formulation (\ref{eq:4}) is preserved. 

\vskip 0.3cm 
After averaging over the gauge fields, the Fourier transform of 
(\ref{eq:gluong}) can be written as 
\be 
G^{\mathrm{gen} \, ab }_{ij}(\vp, p_0) \; = \; \delta ^{ab} \; 
\left(\delta_{ij}-\frac{p_ip_j}{\vp^2}\right) \; 
\frac{F(\vp, p_0)}{  \vp^2 } \; , 
\label{eq:genk}
\en 
where the dimensionless quantity $F(\vp, p_0)$ is a form factor. 
If the residual gauge degree of freedom is not fixed, the 
gauge fields corresponding to different time slices would be 
uncorrelated, and the form factor would not depend on $p_0$. 
We expect that the residual gauge fixing only introduces 
weak correlations between gauge fields of  different time slices 
implying that the form factor only slightly depends 
on $p_0$. A detailed lattice investigation is left to the near 
future. In the present paper, we will focus on the equal time gluon 
propagator (\ref{eq:gluon}). 

\vskip 0.3cm 
In practice, we used a {\it simulated annealing} method 
to locate the minimum of (\ref{eq:gcond}). 
The idea is to consider the gauge functional
$F_U[\Omega]$ as the action of a field theory with respect to the set of
the gauge transformations $\Omega(\vx)$, whose partition function is 
given by 
$$ 
Z_{SA}=\int D\Omega \exp{(-\beta_{SA}F_U[\Omega])}. 
$$
By increasing step by step the free parameter $\beta_{SA}$ 
(which plays the role of the inverse temperature), 
one tries by cooling to retrieve the ground
state of the fictitious field theory.  This procedure corresponds to the
minimization of $F_U[g]$. The thermalization steps can be 
performed using the standard Creutz update algorithm 
supplemented by microcanonical reflection steps. 
After the domain of attraction has been located by simulated 
annealing, we perform {\it iterated overrelaxation} 
to ensure the desired precision of the gauge condition. Details 
will be presented elsewhere. At least in minimal Landau gauge, it 
has turned out that the present method of gauge fixing is 
robust against the impact of Gribov ambiguities~\cite{Bloch:2003sk}.  

\section{ \label{sec:num} Numerical Results }

\subsection{  \label{sec:gluon} Gluon propagator } 

\begin{figure*}
\includegraphics[height=7cm]{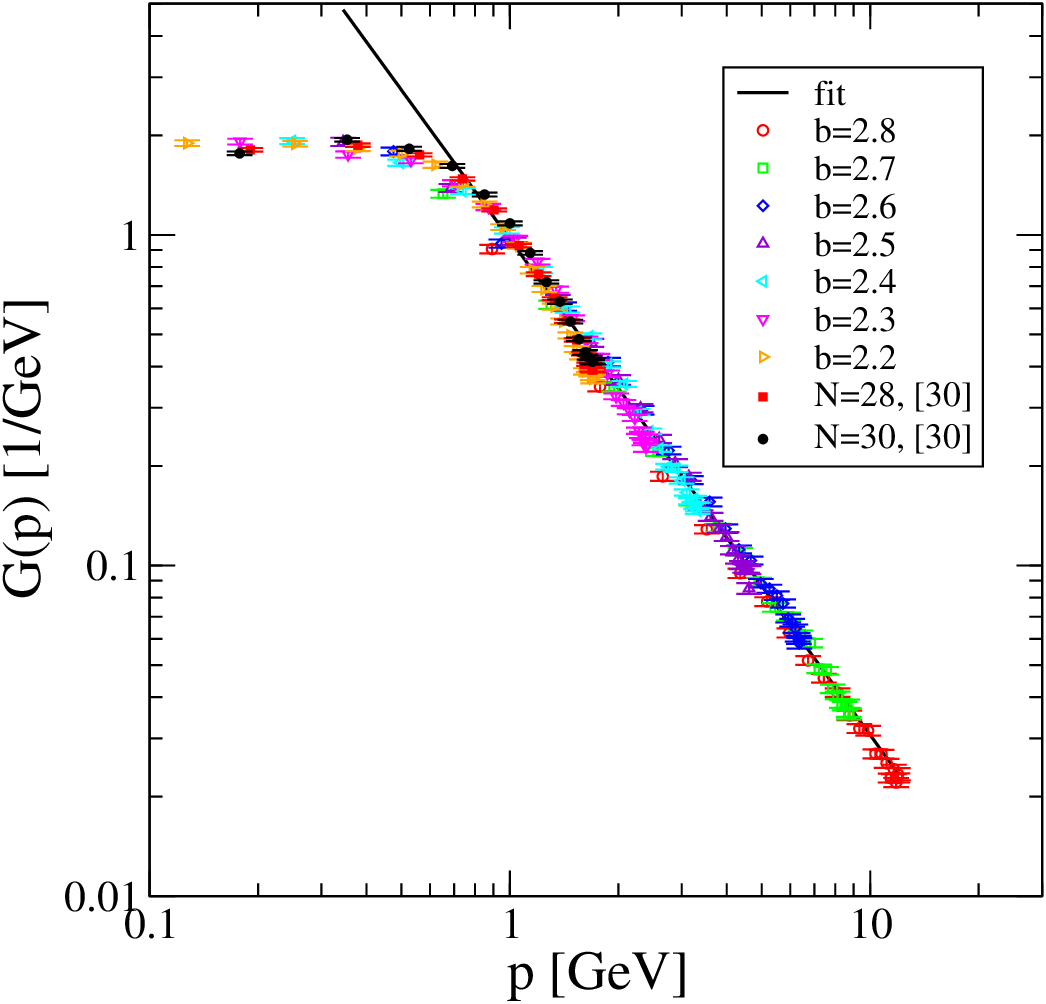} \hspace{0.5cm}
\includegraphics[height=7cm]{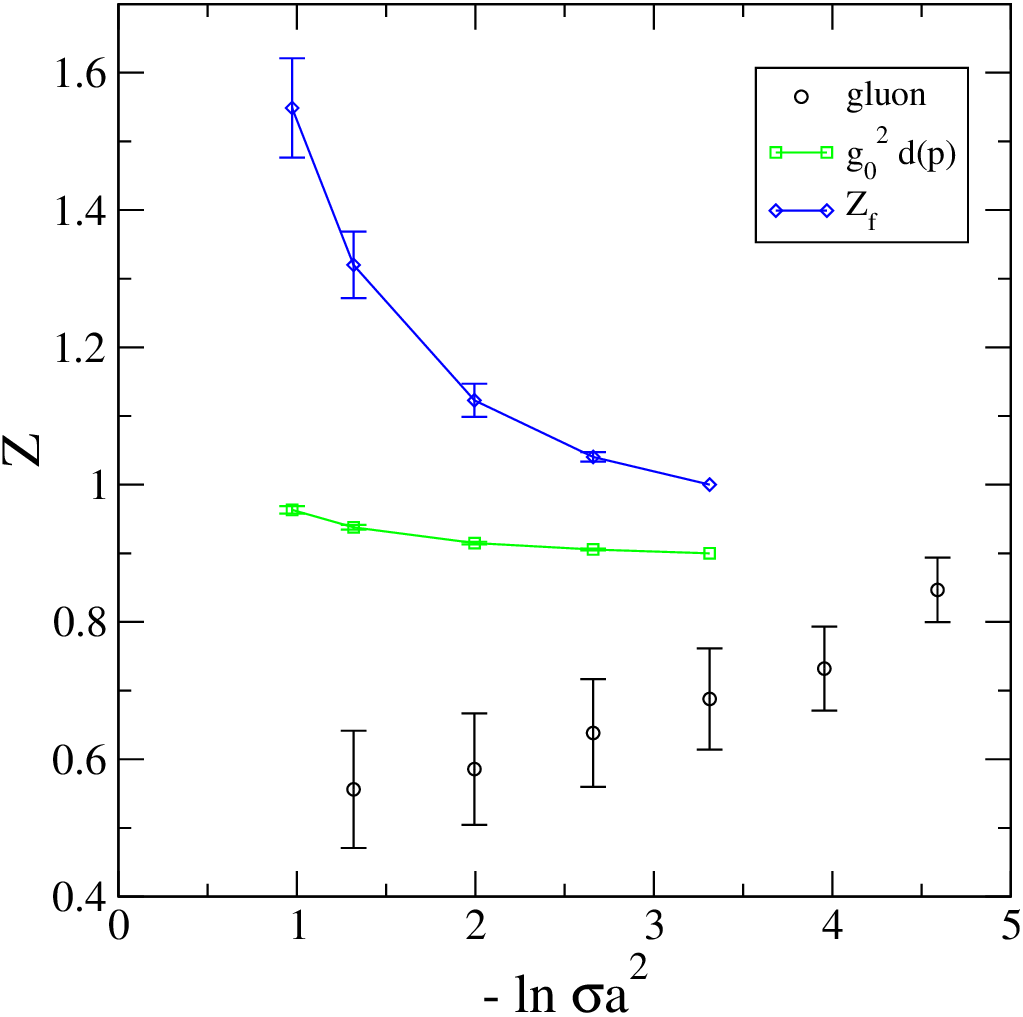}
\vspace{.2cm}
\caption{\label{fig:1} The transverse equal-time gluon propagator 
  as a function of the momentum: our data ($N=42)$ and  data 
  from~\cite{Cucchieri:2000gu} (left panel). The gluon wave function 
  renormalization constant $Z_3$ as a function of the $UV$ cutoff. 
  Also shown are the wave functional renormalization constant 
  of $g_0^2 d(\vp)$ and  $Z_f$ (\ref{eq:vcren}); the y-axis is arbitrarily 
  scaled  (right panel).  } 
\end{figure*}
\begin{figure*}
\vspace{1cm}
\includegraphics[height=7cm]{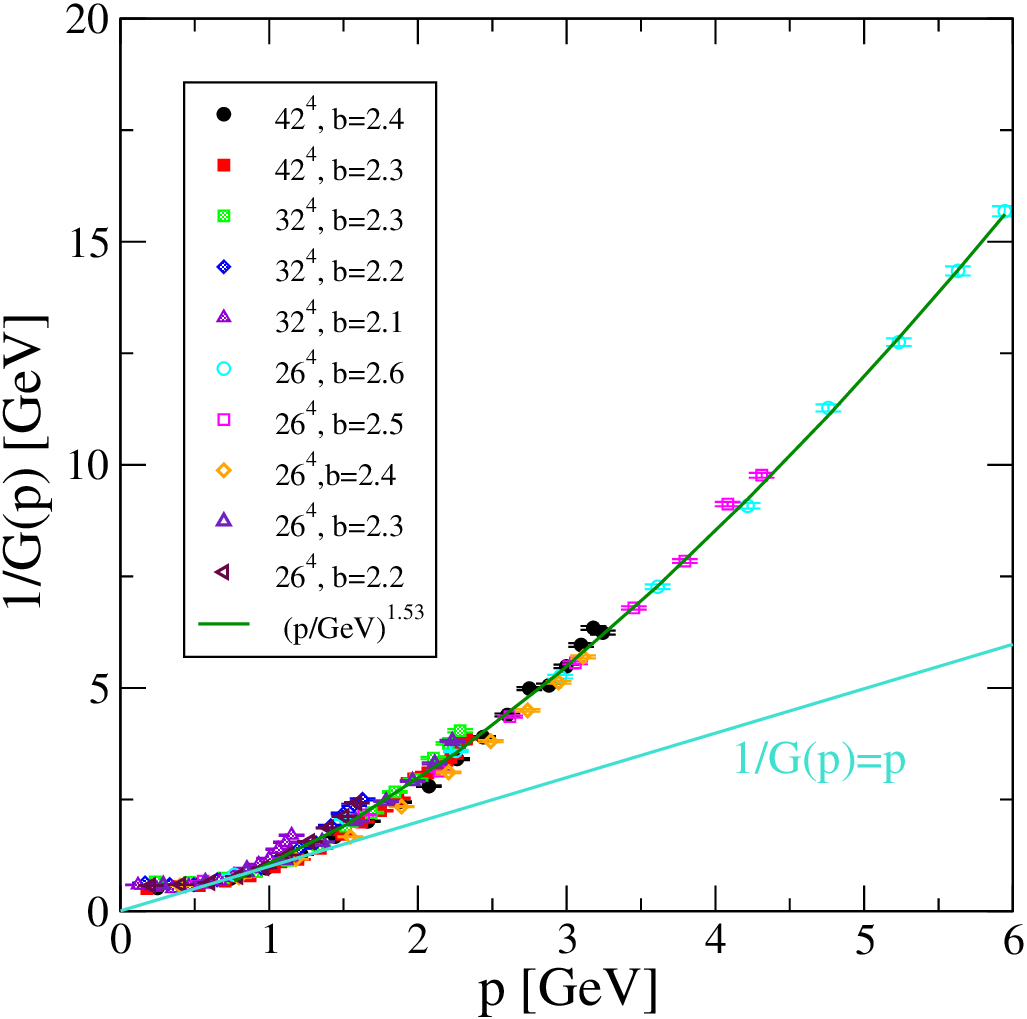} \hspace{0.5cm}
\includegraphics[height=7cm]{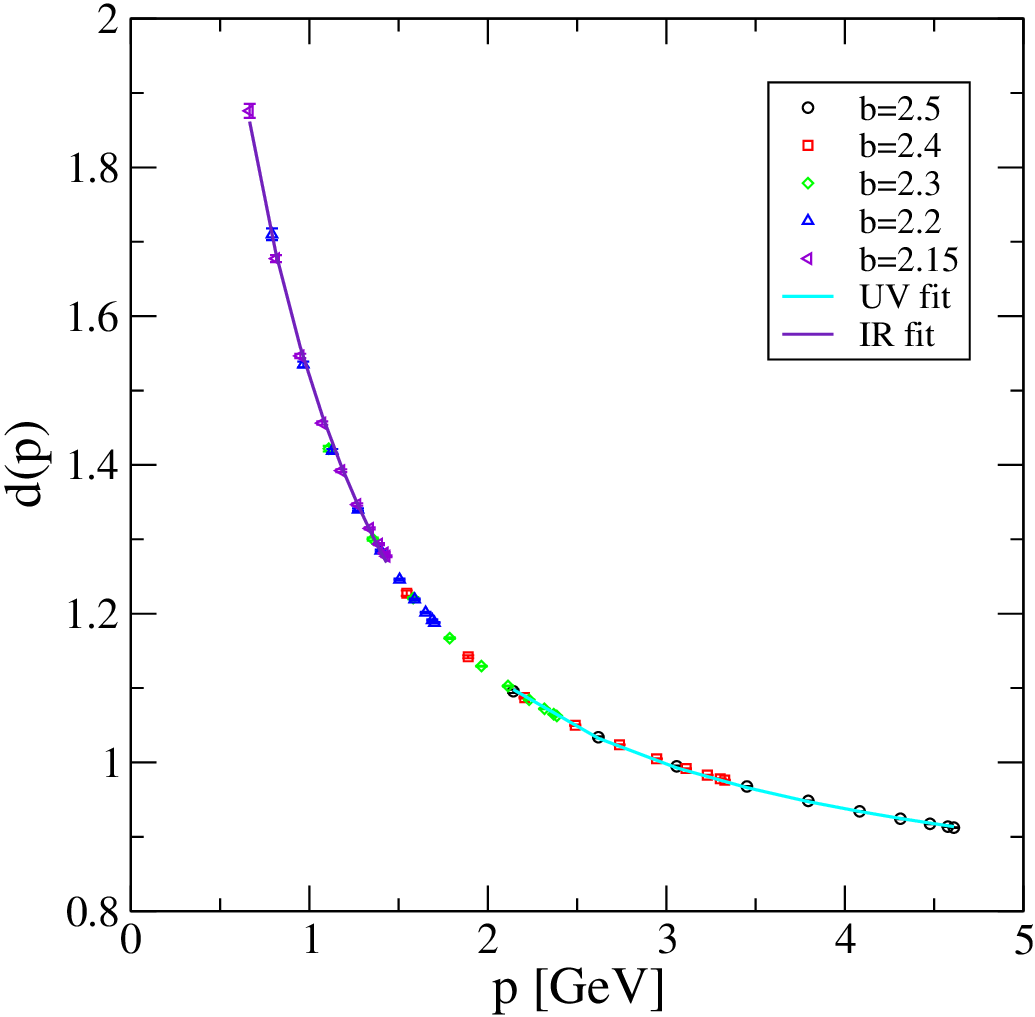} 
\vspace{.2cm}
\caption{\label{fig:2} Momentum dependence of the inverse gluon propagator
  $1/G (\vp)$ (left panel). Ghost form factor (right panel).
  } 
\end{figure*}
Lattice simulations were carried out using $26^4$, $32^4$ and $42^4$ 
lattices. Inverse couplings $\beta \in \{ 2.2, 2.3, 2.4, 2.5, 2.6, 
2.7, 2.8 \}$ 
were employed. Within the momentum window, very good scaling is 
observed implying that cutoff effects (due to the finite value $a$) 
and finite size effects are small. Our final result for the 
transverse equal-time gluon propagator $G(\vp)$ (\ref{eq:gluonm}) 
is shown in figure \ref{fig:1}. Only data for the $42^4$ lattice are 
shown for clarity. We have checked that the data for the 
 $26^4$, $32^4$ lattices fall on top of the same curve. 
As a renormalization condition, we have chosen 
\be 
G(\vp = 1 \, \mathrm{GeV}) \; = \; 1 \; [\mathrm{GeV}]^{-1} 
\en 
for a renormalization point $\mu = 1 \,$GeV. 
For comparison, we have also shown the data obtained by 
Cucchieri and Zwanziger in~\cite{Cucchieri:2000gu}. These data 
are obtained for $\beta =2.2$ and $28^4$, $30^4$ lattices by 
using an iterated overrelaxation method for gauge fixing. 
We find good agreement. 

\vskip 0.3cm 
At small momentum, the propagator becomes roughly momentum independent 
and seems to approach a constant in the IR limit $\vert \vp \vert 
\rightarrow 0$. At large momentum, the momentum dependence is 
well approximated by 
\be 
G(\vp) \propto \frac{1}{\vert \vp \vert } 
\left( \frac{\Lambda _{QCD}}{\vert \vp \vert } \right)^{ \eta } 
\; , \hbo 
\eta = 0.5(1) \; , 
\en 
where $\Lambda _{QCD}$ is the renormalization group invariant scale 
parameter. 
The solid line in figure \ref{fig:1} corresponds to a fit with 
$\eta =0.5$. Also shown are the lattice data from~\cite{Cucchieri:2000gu} 
(by courtesy of Attilio Cucchieri). The large value $\eta $ was 
already anticipated in~\cite{Cucchieri:2000gu}. Since 
the authors focused their investigations on the IR behavior of 
the gluon propagator, they concluded that their data, 
obtained for momenta below $2 \,$GeV, did not reach the perturbative 
regime. In the present paper, the high momentum regime is also 
explored. It turns out that the trend with $\eta $ as large as $0.5$ 
continues at least up to momenta as large as $12 \,$ GeV. 

\vskip 0.3cm 
Also shown in figure \ref{fig:1} is the cutoff dependence 
of the wave function renormalization constant $Z_3$. 
Unfortunately, the achieved numerical precision does not allow for 
a definite conclusion on the functional form which parameterizes the 
cutoff dependence of $Z_3$.

\vskip 0.3cm 
In the context of a variational approach~\cite{Szczepaniak:2001rg,
Szczepaniak:2003ve,Feuchter:2004gb}, $1/G (\vp )$ is 
interpreted as energy dispersion relation of constituent gluon fields. 
Note, however, that the full gluon propagator $G(\vp )$ 
which comprises all non-perturbative effects 
is parameterized by $G(\vp ) = g(\vp )/  \omega _{QP} (\vp ) $, 
where $ g(\vp )$ accounts for effects which are beyond the leading 
order of the quasi-particle (QP) picture. Hence, the interpretation 
of the inverse propagator as dispersion relation must be taken 
with care. Our results for the inverse gluon propagator $ 1/G (\vp ) $ 
for several lattice sizes are shown in figure \ref{fig:2}. 

\begin{figure*}
\vspace{1cm}
\includegraphics[height=7cm]{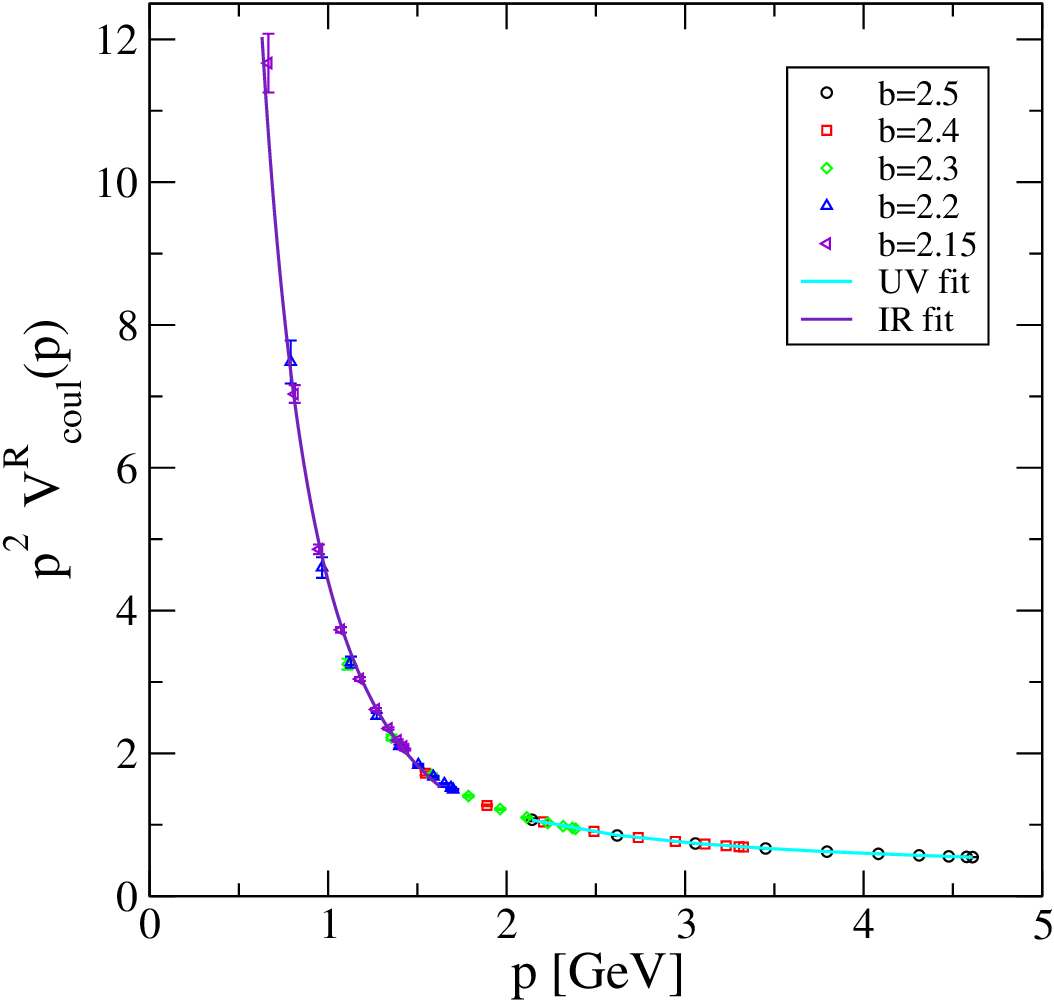} \hspace{0.5cm}
\includegraphics[height=7cm]{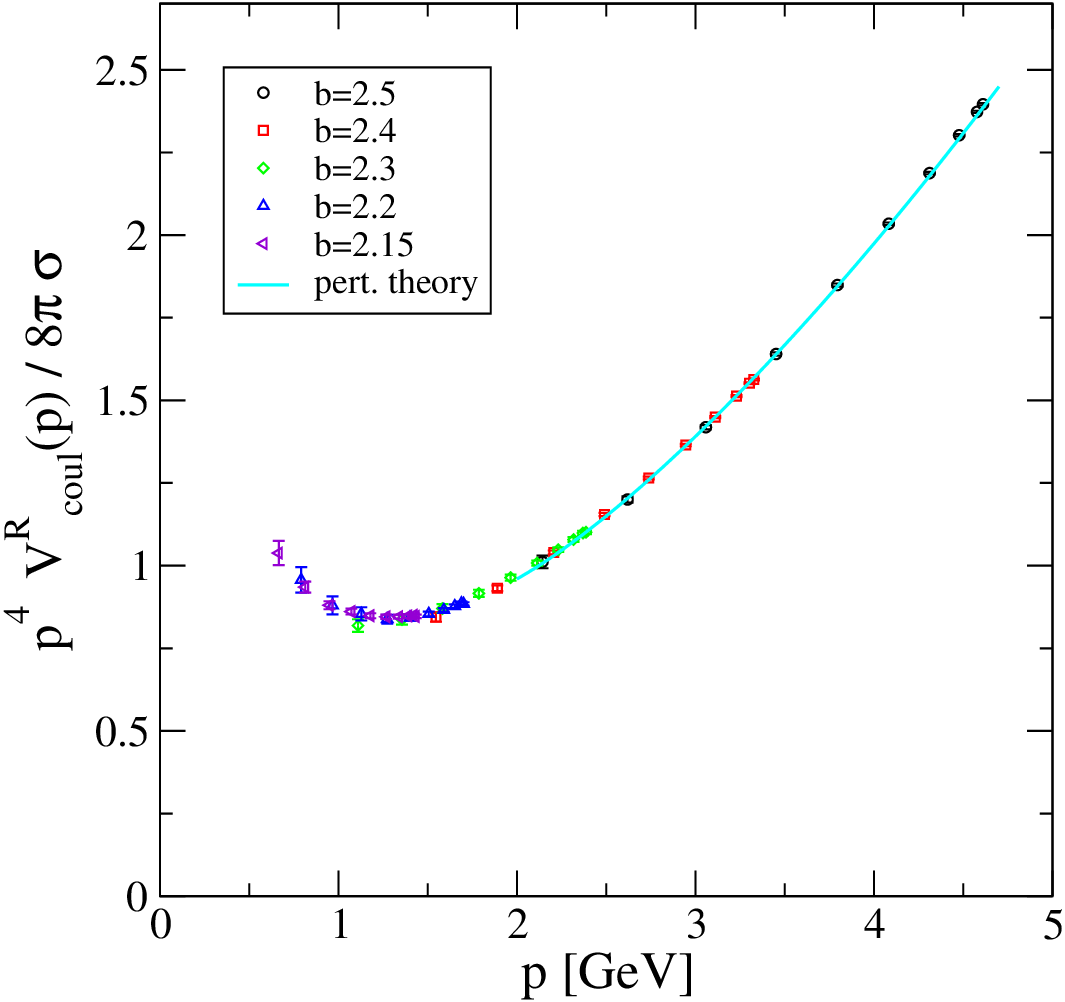} 
\vspace{.2cm}
\caption{\label{fig:3} $p^2 \; V _{\rm Coul}(\vp ) $ 
  (Coulomb potential) as a function of the momentum $\vp$ (left panel). 
   $p^4 \; V _{\rm Coul}(\vp )/8\pi $  in units of the string tension 
   $\sigma $ (right panel). 
  } 
\end{figure*}
\subsection{ Ghost form factor } 

In order to obtain the ghost propagator (\ref{eq:ghostp}), 
an inversion of the Faddeev-Popov matrix is performed after 
(possible) zero modes have been removed~\cite{Suman:1995zg}. 
This procedure requires the solution of large, but sparse, systems 
of equations. In order to remove zero mode contributions, we solve 
$$ 
M \; M \; x \; = \; M \; b 
$$
rather than the system $M x = b$ (see the method 
proposed in~\cite{Suman:1995zg}). For this purpose, we use the 
{\tt minres} algorithm with Jacobi preconditioning~\cite{paige}. 
Our final result for the ghost form factor $d(\vp )$ 
obtained from simulations using a $26^4$ lattice and 
$\beta \in \{2.15, 2.2, 2.3, 2.4, 2.5 \}$ is shown in 
figure \ref{fig:2}, right 
panel. We observe perfect scaling. To be specific, we choose 
$$ 
d(\vp = 3 \, \mathrm{GeV}) \; = \; 1  
$$
as a renormalization condition. 

\vskip 0.3cm 
In order to explore the high momentum dependence, we make 
a logarithmic ansatz supplemented with an anomalous dimension 
$\gamma _g$, i.e., 
\be 
d(\vp ) \; = \; \frac{a_{uv}}{ \ln \Bigl( \vert \vp \vert / \Lambda _{QCD} 
\Bigr) ^{\gamma _{go}} } \; , \hbo p \gg  \Lambda _{QCD} \; . 
\en 
The UV fit shown in figure \ref{fig:2} (right panel) corresponds to 
$$ 
a_{uv}\; \approx \; 1.03(2) \; , \hbo \gamma _{go} \; = \; 0.26(2) \; . 
$$
Since the fitting parameters are strongly correlated, it is difficult 
to extract a reliable value for $\Lambda _{QCD}$. Taking also into 
account the high momentum behavior of the Coulomb potential 
(see below), we find that 
$$
\Lambda _{QCD} \; = \; 0.96(5) \, \mathrm{GeV} 
$$ 
reproduces the UV data. 
Such a high value for the Yang-Mills scale parameter seems to 
be generic for the lattice regularization (see~\cite{Bloch:2003sk}). 
For the IR analysis, we adopt a simple scaling law: 
\be 
d(\vp ) \; = \; \frac{a_{ir}}{ \Bigl( \vert \vp \vert ^2 / \Lambda ^2_{QCD} 
\Bigr)^\kappa } \; , \hbo p \ll \Lambda _{QCD}  \; . 
\en 
The IR fit is also shown in  figure \ref{fig:2} (right panel). We find 
\be 
a_{ir} \; \approx \; 1.55(1) \; , \hbo \kappa = 0.245(5) \; . 
\en 
Hence, our lattice results suggest that the ghost form factor 
roughly diverges like $1/\sqrt{\vert \vp \vert } $ in the IR limit 
$\vert \vp \vert \rightarrow 0$. 

\vskip 0.3cm 
Let us explore the ghost wave function renormalization constant 
$\widetilde{Z}_3$. If one applies the momentum matching technique 
to $g_0^2 \, d_0(\vp, \beta ) $, one finds that the matching factors 
are constant within the numerical precision, see 
figure \ref{fig:1}, right panel. Our findings therefore 
suggest that the bare coupling squared times the bare ghost propagator, 
i.e., 
\be 
g_0^2 \; D_0^{ab}(\vp) \; , 
\en 
is a renormalization group invariant. 

\subsection{Coulomb potential }

In the present approach, the Coulomb potential appears as 
a convolution of Green functions (see (\ref{lat:vcoul})). In full 
Yang-Mills theory, wave functional (and vertex) renormalization applies 
(see (\ref{eq:vcren})), and the prefactor of the potential must be 
specified by a renormalization condition. The prefactor can be 
determined by demanding that the perturbative result 
is recovered~\cite{Cucchieri:2000hv} at large momentum, i.e., 
\be 
\vp^2 \; V _{\rm Coul}(\vp) \; \approx \; 
\frac{ 6\pi }{ 11 \; \ln \vert  \vp \vert ^2  / \Lambda ^2_{QCD} } 
\label{eq:high}
\en 
for $\vert \vp \vert \gg  \Lambda _{QCD}$. 
An elegant method which circumvents the cumbersome determination 
of the prefactor was put forward in~\cite{Greensite:2003xf,Greensite:2004ke}. 

\vskip 0.3cm 
Our numerical findings for $p^2 \; V _{\rm Coul}(\vp) $ 
are summarized in figure \ref{fig:3} (left panel). At large momentum, 
the lattice data nicely show the logarithmic correction. 
The data of low energy regime are well reproduced by the scaling 
ansatz 
\be 
\vp^2 \; V _{\rm Coul}(\vp) \; \approx \; 
\frac{ c }{ ( \vert  \vp \vert^2 / \Lambda ^2_{QCD})^\delta } 
\; , 
\en 
where $ c $ is related to the Coulomb string tension $\sigma _{\rm Coul}$ 
by $\sigma _{\rm Coul} \, = \, c \, \Lambda ^2_{QCD} / 8\pi $ 
in case $ V _{\rm Coul}$ is compatible with linear 
confinement ($\delta =1$). 
Here, we find (see figure \ref{fig:2}, left panel) 
\be 
\delta \; = \; 1.05(5) \; , \hbo c \; = \; 4.9(2) \; . 
\en 
Also shown in figure \ref{fig:2} (right panel) is the combination 
$$
\vp^4 \; V _{\rm Coul}(\vp) \; / \; 8 \pi \sigma 
$$ 
which should approach $\sigma _{\rm coul} / \sigma  $ 
in the limit $\vert \vp \vert \rightarrow 0$.  
Already for $p>2 \,$GeV, the perturbative result is recovered to 
good precision. A plateau is reached at the intermediate momentum 
range $1 \, \mathrm{GeV} < p < 2  \, \mathrm{GeV}$. 
The function is slightly increasing again for $p<1 \, \mathrm{GeV}$. 
The data within the observed momentum window are stable 
against finite size effects. An extrapolation $p \rightarrow 0$ 
is cumbersome. Our findings suggest that the inequality $\sigma \le \sigma  
_{\rm coul} $ is almost saturated. This conclusion is in agreement with 
the results published in~\cite{Cucchieri:2002su}. 
However, we point out that values $ \sigma _{coul} 
/ \sigma $ ranging from $2$ to $3$ reported 
in~\cite{Greensite:2003xf,Greensite:2004ke} cannot be ruled out from 
the present data. Larger lattices and higher statistics will be 
necessary to explore the deep infrared regime.

\subsection{Factorization } 

\begin{figure}
\vspace{1cm}
\includegraphics[height=7cm]{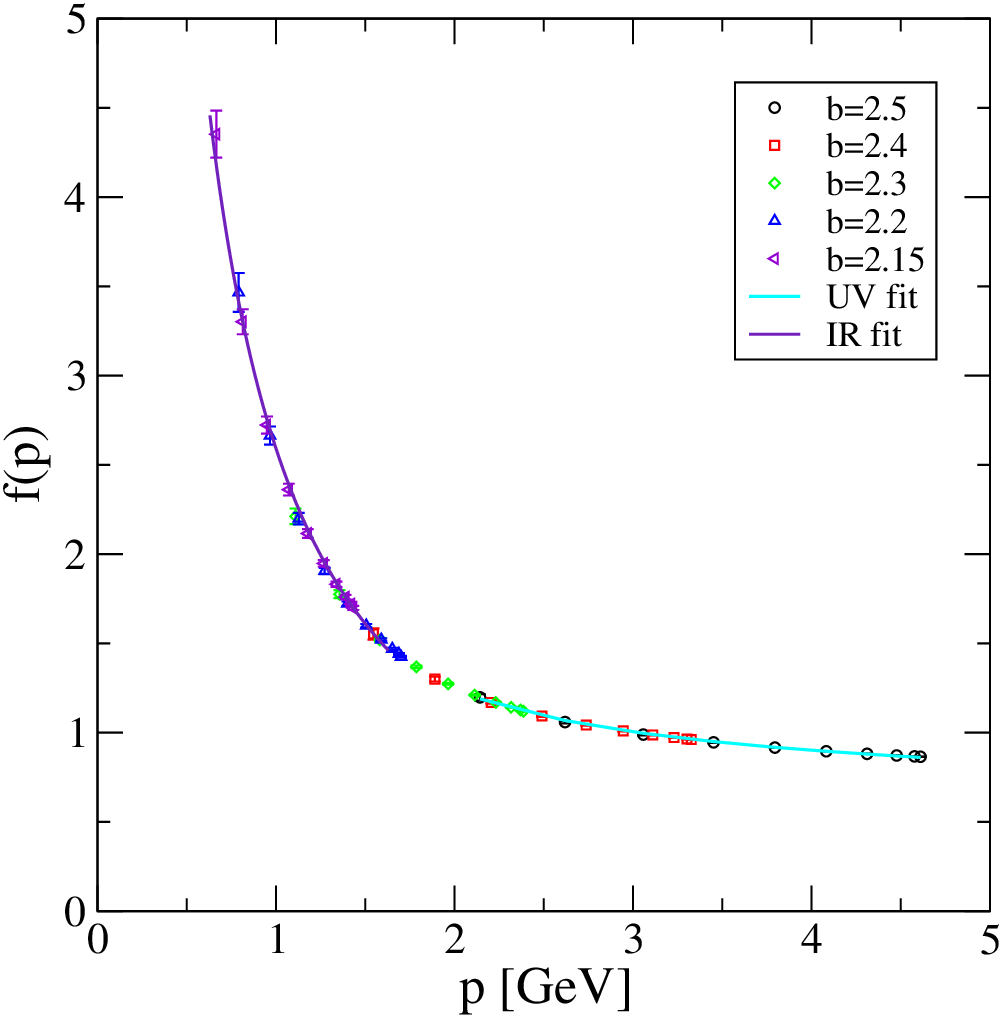} \hspace{0.5cm}
\vspace{.2cm}
\caption{\label{fig:4} 
  The factorization function $f(\vp)$.  
  } 
\end{figure}
From the results of the previous subsections it is already clear 
that the function $f(\vp)$ cannot weakly depend on $
\vert \vp \vert $. Factorization such as (\ref{eq:fa}) is generic 
if the fields are distributed according to a Gaussian probability 
distribution. Since the ghost fields are strongly interacting
at low momenta, one expects that $f(\vp)$ significantly 
depends on the momentum in the IR regime. 

\vskip 0.3cm 
In order to extract  $f(\vp)$ from the lattice data we applied 
the following procedure: Dividing the bare data for 
$\vp ^2 V^{0}_{coul}$ with the bare ghost form factor squared 
provides information on the unrenormalized function $f^0(\vp)$. 
Applying the momentum matching technique~\cite{Bloch:2003sk} 
yields the renormalization constant $Z_f$ and the renormalized function 
$f(\vp )$. Our numerical result for $Z_f $ 
as a function of the UV cutoff is consistent with (see figure \ref{fig:1}, 
right panel) 
$$ 
\lim _{\beta \to \infty } Z_f (\beta , \mu ) \; = \; \mathrm{constant} 
\; . 
$$ 
The function $f(\vp)$ is shown in figure \ref{fig:4}. 
We observe a weak momentum dependence in the UV regime, and 
an IR singularity at small momentum. The high energy data 
are reproduced by the ansatz 
\be 
f(\vp ) \; = \; \frac{a^f_{uv}}{ \ln \Bigl( \vert \vp \vert / \Lambda _{QCD} 
\Bigr) ^{\gamma _{f}} } \; , \hbo p \gg \Lambda _{QCD} \; . 
\en 
The UV fit shown in figure \ref{fig:4} corresponds to 
$$ 
a^f_{uv}\; \approx \; 1.05(4) \; , \hbo \gamma _{f} \; = \; 0.47(3) \; . 
$$
The low momentum data support the existence of a $1/\vert \vp \vert $
singularity, i.e., 
\be 
f(\vp ) \; = \; \frac{a^f_{ir}}{ \Bigl( \vert \vp \vert^2 /  
\Lambda ^2_{QCD} \Bigr)^{\kappa _f} } \; , \hbo p <  \Lambda _{QCD} \; . 
\en 
with 
\be 
a^f_{ir} \; \approx \; 2.7(1) \; , \hbo \kappa _f = 0.58(5) \; . 
\en 
Our findings suggest the following interpretation: due to asymptotic freedom, 
the expectation value on the left hand side of (\ref{eq:fa}) 
factorizes in the high momentum regime. At low momentum, non-Gaussian 
correlations between the ghost fields play an important role and 
the factorization assumption is ruled out. 

\vskip 0.3cm 
Finally, we point out that the sum rule, dictated by perturbation 
theory, i.e., $2 \, \gamma _{go} + \gamma _f =1 $, is satisfied to 
good precision: 
\be 
2 \; \gamma _{go} + \gamma _f \; = \; 0.99(12) \; . 
\en 
This serves as a consistency check of the lattice renormalization 
procedure. 

\section{Critical remarks  } 

\subsection{ Equal time gluon propagator } 
In subsection \ref{sec:gluon}, the lattice data for the transverse 
equal time gluon propagator $G(\vp)$ are consistent with the 
high momentum behavior $G(\vp ) \propto \vert \vp \vert ^{-1 - \eta }$, 
$\eta = 0.5(1)$. 
It is tempting to conclude that asymptotic freedom implies the 
decrease of the equal-time gluon propagator $G(\vp)$ like 
$1/\vert \vp \vert $ for large momenta. We stress that this conclusion 
does not necessarily apply: 
The equal-time gluon propagator $G(\vp)$ is derived from the 
generalized propagator (\ref{eq:genk}) by momentum projection 
i.e., 
\bea  
G(\vp) &\propto &  Z_3^{-1} \int dp_0 \; G^\mathrm{gen}(\vp, p_0) 
\nonumber \\ 
&=&  Z_3^{-1} \int dp_0 \; \frac{F(\vp, p_0)}{ \vp^2 } \; . 
\ena
After  proper renormalization, the dimensionless form factor 
can be parameterized by 
\be 
F(\vp, p_0) \; = \; h \left( \frac{\Lambda _{QCD}}{\vert \vp \vert }, 
 \frac{p_0}{\vert \vp \vert } \right) \; , 
\en 
where $\Lambda _{QCD}$ is the low energy scale parameter (e.g., 
given by the string tension). For large momenta 
$\vert \vp \vert \gg \Lambda _{QCD}$, one obtains 
\be 
G(\vp) \approx  Z_3^{-1} \frac{1}{\vert \vp \vert } \int du \; 
h(0,u) \; . 
\en 
If the function $h(x,y)$ is 
completely regular and bounded for $x, y \in [0, \infty [$, the integral 
in the latter equation would exist implying that 
$G(\vp) \propto \frac{1}{\vert \vp \vert } $ for large $\vert \vp \vert $. 
However, the form factor $F(\vp, p_0)$ weakly depends on $p_0$ 
by the choice of the gauge (see discussions below (\ref{eq:genk})). 
Let us speculate that for this reason we might assume for $u\gg 1 $ that 
$$
h(0,u) \propto \frac{1}{ u ^{1-\eta } } \; \hbo 0< \eta < 1 \; . 
$$ 
In this case, the asymptotic form of the equal time propagator would be 
given by 
$$ 
G(\vp) \; \propto \; \frac{1}{\vert \vp \vert } \left( 
\frac{ \Lambda _{QCD} }{ \vert \vp \vert } \right)^\eta 
$$ 
where the divergent factor $(\Lambda _{UV}/\Lambda _{QCD})^\eta $ 
has been absorbed by the wave function renormalization 
constant $Z_3$. 
The above picture is only one scenario which would explain the 
lattice data. An lattice investigation of the $p_0$, $\vp$ 
dependence of the generalized gluon propagator 
$G^{\mathrm{gen} \, ab }_{ij}(\vp, p_0) $ in (\ref{eq:genk}) 
will reveal whether the above scenario applies for 
Coulomb gauged YM-theory. Such an investigation is left to 
future work. 

\subsection{Renormalization group } 
Perturbation theory in Coulomb gauge~\cite{Zwanziger:ez,Baulieu:1998kx} 
is plagued by superficial divergences originating from 
instantaneous loops. When ``supplementary rules'' are introduced 
to define zero momentum integrals, a self-consistent framework 
is obtained. Within this approach, the $Z_g$ and the ghost 
renormalization constants, $Z_C$ and $Z_{\bar{C}}$, are related 
by~\cite{Zwanziger:ez} 
\be 
Z_g \; Z_C \; = \; 1 \; , \hbo 
Z_g^2 \; Z_C \; Z_{\bar{C}} \; = \; 1 \; - \; \frac{7 \, g^2 }{ 
12 \pi ^2 \; \epsilon } \; . 
\label{eq:per}
\en 
We point out that our lattice results indicate that the product 
of bare gauge coupling $g_0$ and bare ghost form factor $D_0$, 
i.e., $g_0^2 \, D_0$, is finite, which contradicts (\ref{eq:per}). 
Further studies, involving simulation with larger lattices on one hand and 
partial resummations of perturbation theory to handle 
instantaneous loops on the other hand, seem necessary to resolve the above 
discrepancy.

\section{Conclusions } 

A thorough lattice study of the equal-time propagators of 
four dimensional SU(2) Yang-Mills theory in Coulomb gauge has been 
performed. Our simulations extend previous studies to larger 
lattices and number of different lattice spacings. 
Our lattice data are in good agreement with those of earlier 
publications~\cite{Cucchieri:2000gu,Greensite:2003xf,Greensite:2004ke}. 

\vskip 0.3cm 
We studied the transverse (equal time) gluon propagator $G(\vp)$, 
the ghost form factor $d(\vp )$ and the Coulomb potential 
$V_{\rm Coul}(\vp)$ (\ref{eq:coul}). 
With the help of the momentum matching technique~\cite{Bloch:2003sk}, 
the wave functional renormalization constants are obtained for the 
gluon and ghost fields. In particular, we find that 
the combination  $g_0^2 \; d_0(\vp )$ of the bare 
gauge coupling $g_0$ and the bare ghost form factor $d_0$ 
is renormalization group invariant. 

Let us focus here on the high and the low momentum behavior. 
For $\vert \vp \vert \gg 1 \, $GeV, our results for the (equal time) 
gluon propagator show a large anomalous dimension, i.e., 
$$
G (\vp ) \propto 1 / \vert \vp \vert ^{1+ \eta } \; , \hbo 
\eta \approx 0.5(1) \; . 
$$ 
The high momentum behavior of the ghost form factor and 
the factorization function $f(\vp )$ show the characteristic 
logarithmic momentum dependence, i.e., 
\bea 
d(\vp ) &\propto & \ln \left( \frac{ \vert \vp \vert }{ \Lambda _{QCD} } 
\right) ^{- \gamma _{go}  } \; , \hbo \gamma _{go} \; = \; 0.26(2) \; , 
\nonumber \\ 
f(\vp ) &\propto & \ln \left( \frac{ \vert \vp \vert }{ \Lambda _{QCD} } 
\right) ^{- \gamma _{f}  } \; , \; \; \hbo \gamma _{f} \; = \; 0.47(3) \; , 
\nonumber 
\ena 
while a scale parameter of $ \Lambda _{QCD} = 0.96(5)$\, GeV is consistent 
with the lattice data. The high momentum behavior of the 
Coulomb potential known from perturbation 
theory~\cite{Cucchieri:2000hv}, i.e., 
$$
\vp^2 \; V _{\rm Coul}(\vp) \; \approx \; 
\frac{ 6\pi }{11 \; \ln \vert  \vp \vert ^2  / \Lambda ^2_{QCD} } 
$$
is recovered to high precision. The anomalous dimensions extracted 
from the lattice data are consistent with the sum rule 
$2 \, \gamma _{go} + \gamma _{f}=1$. 

\vskip 0.3cm 
In the IR limit $\vert \vp \vert \rightarrow 0$, the quantities 
$d(\vp)$ and $f(\vp)$ 
develop singularities. The data for the Coulomb potential are 
parameterized in the IR regime by 
$$
V _{\rm Coul}(\vp) \; \approx \; 
\frac{8 \pi \, \sigma _{coul} }{ \vert  \vp \vert ^4 } \; , 
$$ 
and are consistent with linear confinement. 
We find that the inequality 
$$ 
 \sigma _{coul} \; \le \;  \; \sigma \; 
$$
is almost saturated in agreement with~\cite{Cucchieri:2002su}. 
We stress, however, that values in the range $ \sigma _{coul} \; 
= \; (2-3) \,  \sigma $ as reported 
in~\cite{Greensite:2003xf,Greensite:2004ke} are not ruled out 
by our data. 
The singularity of $V _{\rm Coul}(\vp) \propto 1/\vert  \vp \vert ^4 $, 
responsible for 
linear confinement, arises from IR singularities of the functions  
$d(\vp )$ and $f(\vp)$: 
$$ 
d(\vp ) \;\propto \; \frac{ 1 } { \vert \vp \vert ^{0.49(1) }} \; , 
\hbo 
f(\vp ) \;\propto \;  \frac{ 1 } { \vert \vp \vert ^{1.17(10) } } \; . 
$$
The complete momentum dependence for the above functions can be found 
in section \ref{sec:num}.

\vspace{.3cm}
{\bf Acknowledgments.} 
We thank R.~Alkofer, J.~Greensite, B.~Gr\"uter, S.~Olejnik, H.~Reinhardt 
and D.~Zwanziger for helpful discussions. 
 LM is supported by the European Graduate School 
"Hadronen im Vakuum, in Kernen und Sternen" under contract DFG GRK683.


\end{document}